 \documentclass[12pt,preprint]{aastex}

\usepackage{amsmath}
 \usepackage{color}

\shorttitle{Bremsstrahlung for X-ray Thermal Component of GRB 130925A}

\shortauthors{Liu \& Mao}

\begin{document}


\title{Is Bremsstrahlung A Possible Mechanism to Explain the Thermal Feature in the GRB 130925A X-ray Afterglow?}


\author{Jie-Ying Liu\altaffilmark{1,2,3} and Jirong Mao\altaffilmark{1,2,3}
}

\altaffiltext{1}{Yunnan Observatories, Chinese Academy of Sciences, Kunming 650011, Yunnan Province, China}
\altaffiltext{2}{Center for Astronomical Mega-Science, Chinese Academy of Sciences, 20A Datun Road, Chaoyang District, Beijing 100012, China}
\altaffiltext{3}{Key Laboratory for the Structure and Evolution of Celestial Objects, Chinese Academy of Sciences, Kunming 650011, China}

\email{jirongmao@mail.ynao.ac.cn}

\begin{abstract}
It has been reported that some X-ray spectra of gamma-ray burst (GRB) afterglows cannot be fitted by a simple power law. A blackbody component is added to precisely fit the thermal feature in these spectra. Alternatively, we propose that bremsstrahlung radiation can also be one possible mechanism to explain the thermal component of the GRB X-ray afterglow. In particular, we examine the X-ray afterglow of the ultra-long GRB 130925A in this paper. By our calculation, we find that the X-ray thermal component observed by both {\it Swift}-XRT and {\it NuSTAR} can be well explained by the bremsstrahlung radiation. Our results indicate that the GRBs with the bremsstrahlung emission in the X-ray afterglow could be born in a metal-rich and dusty environment.
\end{abstract}




\section{Introduction}
It is widely accepted that gamma-ray burst (GRB) physics can be well described by the fireball model (e.g., Piran 1999). Although synchrotron radiation is often adopted to explain a power-law feature in a general GRB afterglow spectrum, a thermal component fitted by a blackbody shape has been identified in some {\it Swift}-XRT observations \citep{sparre12,starling12,valan18}.

GRB 130925A was triggered by {\it INTEGRAL} and {\it Fermi}-GBM. It was also detected by Konus-{\it Wind} and {\it Swift}. The redshift of this burst is $z=0.348$ \citep{sudi13}. It is an ultra-long GRB with the first episode at a duration of about 900 s and the second episode at the time of $2-3$ ks after the trigger. The total duration of the prompt emission is about 20 ks. Multiple flares are shown in the X-ray light curve \citep{evans14}. This burst was also observed by {\it NuSTAR} \citep{bellm14}. Some optical/infrared flares were detected \citep{greiner14}. The radio emission at 7 GHz had a peak flux only 2.2 days after the trigger \citep{horesh15}. It is important to note that a detailed spectral analysis on the thermal emission with the combined data of {\it Swift}-XRT and {\it NuSTAR} was given by \citet{basak15b}.

Some interesting properties of GRB 130925A have been presented. First, in order to explain the long duration of some ultra-long GRBs, blue supergiant (BSG) as a special GRB progenitor class has been proposed \citep{gendre13,nakauchi13,stratta13}. \citet{gendre13} also mentioned a red supergiant (RSG) origin for ultra-long GRBs. From a statistic point of view, \citet{levan14} and \citet{boer15} found that ultra-long GRBs are different from normal GRBs. If GRB 130925A has a BSG/RSG origin, the radiation regions of both the prompt emission and the afterglow can be at very large fireball radii.
Second, the host galaxy observation of GRB 130925A provides an optical extinction of $A_V=2.2$, and $A_V=7.7$ was derived from the {\it Swift}-XRT observation \citep{evans14}. The detection of \citet{schady15} indicates that the host galaxy has supersolar metallicity and the line-of-sight extinction is $A_V=5.0$. These results imply that the surrounding medium of GRB 130925A may provide more ionized particles than that of normal GRBs.
Third, the standard external shock model is hard to use to explain the low-luminosity X-ray afterglow of GRB 130925A, and an additional component on the dust-echo scattering to the prompt emission is suggested \citep{evans14}. A detailed analysis of the dust scattering for GRB 130925A was given by \citet{zhao14}.

If the external shock model is not unique for the complete explanation of the GRB 130925A X-ray afterglow, we may consider other radiation mechanisms besides the synchrotron radiation.
In this paper, we focus on the X-ray thermal component. One may usually adopt the blackbody spectrum to explain the thermal emission. Alternatively, we attempt to use bremsstrahlung radiation to investigate the thermal emission of the GRB 130925A X-ray afterglow. We concentrate on three issues related to this topic. First, the bremsstrahlung temperature should be consistent with the thermal component temperature in the X-ray afterglow of GRB 130925A. The modeling temperature can be constrained by the observed spectrum. Second, the emissivity of the bremsstrahlung radiation is proportional to $Z^2n_en_i$, where $n_e$ is the electron number density, $n_i$ is the ion number density, and $Z$ is the atomic number \citep{rybicki1979}. This means that heavy elements can be ionized due to high temperature in GRB X-ray afterglow and have an important contribution to the bremsstrahlung radiation. If GRB 130925A occurred in a dusty and supersolar metallicity environment as mentioned above, there can be enough ionized particles to produce the bremsstrahlung radiation. Third, the combined data from both {\it Swift}-XRT and {\it NuSTAR} observations provide us a relatively wide energy range to fully illustrate the thermal component properties in the X-ray spectrum. In particular, we investigate one possible bremsstrahlung feature above 10 keV with the {\it NuSTAR} data reduced by Basak \& Rao (2015).

We describe the calculation processes and compare our results to the observational data in Section 2. Discussion and conclusions are given in Section 3. We use the standard cosmological parameters as: $H_{0}=72~\rm{km~s^{-1}~Mpc^{-1}}$, $\Omega_{\Lambda}=0.7$, and $\Omega_{M}=0.3$.

\section{Bremsstrahlung Radiation For X-Ray afterglow of GRB 130925A}
We follow the bremsstrahlung calculation presented by Rybicki \& Lightman (1979). We write the emissivity of the bremsstrahlung cooling as
\begin{equation}\label{e_ff}
e_{\nu}(\nu)=6.8\times10^{-38}Z^2n_en_iT^{-1/2}\rm{exp}(-h\nu/kT )\bar {g}_{\rm ff}\,~{\rm  erg\,s^{-1}\,cm^{-3}\,Hz^{-1}},
\end{equation}
and it can be transferred to the observer frame with the temperature $T_{\rm{obs}}=\Gamma T/(1+z)$, where $Z$ is the atomic number, and $\bar{g}_{\rm ff}$ is the Gaunt factor. Here, $n_i$ and $n_e$ are the ion and electron number densities, respectively, and we assume $n_i=n_e$ and $Z=1$. The forward shock sweeping the surrounding medium can enhance the number density to be $\Gamma^2n_i$ and $\Gamma^2n_e$, and we also assume that the radiation is inside the shell with a thickness of $\Delta H\sim R/\Gamma^2$ compressed by the bulk Lorentz factor in the observer frame, where $R$ is the fireball radius \citep{blandford76}. Thus, the radiation region has the volume of $V_b=4\pi R^2\Delta H$ in the observer frame. Finally, the observed bremsstrahlung flux is given by
\begin{equation}\label{eq-flux}
\nu F_\nu=\frac{V_b \nu e_{\nu}(\nu)}{4\pi D_{\rm L}^2},
\end{equation}
where $\nu e_{\nu}(\nu)$ is the observed emission with a unit of $\rm{erg~s^{-1}~cm^{-3}}$,
the luminosity is given by $V_b\nu e_{\nu}(\nu)$, and $D_{\rm L}$ is the redshift-dependent luminosity distance.

GRB 130925A was observed by {\it Swift}-XRT and {\it NuSTAR} in the wide energy band of $0.3-80$ keV. The combined observational data provide a good opportunity to investigate the GRB thermal component. We take the observational data analyzed by \citet{basak15b}.
The observational epoch is 1.8 days after the {\it Swift}-BAT trigger,
and the redshift-dependent distance is $D_{\rm L }=5.7\times10^{27}\,\rm cm$ (Evans et al. 2014).
We use the bremsstrahlung radiation to investigate the thermal feature of the GRB 130925A X-ray emission. It seems that the thermal feature dominated by the bremsstrahlung radiation plus a nonthermal power-law emission with a slope of 1.7 given by the synchrotron radiation can fit the observational data. GRB afterglow in the X-ray band has a bulk Lorentz factor of about 3-10 as a result of shock deceleration by dense circumstellar medium in general (e.g., M\'{e}sz\'{a}ros et al. 1998). The exact number of the Lorentz factor is dependent on not only the ejection structure and the GRB environment, but also the magnetic energy and the electron energy \citep{mao01}. Here, we assume that the bulk Lorentz factor is 10 at the observational time of 1.8 days after the GRB trigger. The fireball radius can be estimated as $R\sim 4\Gamma^2ct\sim 10^{18}$ cm (e.g., Sari 1997). When we choose the electron temperature of 50.0 keV in the observer frame and the particle number density of $3.0\times 10^4~\rm{cm^{-3}}$, we can reproduce the thermal component of the X-ray spectrum in GRB 130925A. We take the above numbers as the reference values.

The calculation results are shown in Figure 1. We examine the effect of the different bremsstrahlung temperatures in the top left panel. The effect of the different particle number densities is shown in the top right panel. We can see the effect of the different bulk Lorentz factor numbers in the bottom left panel. We obtain the different effects by both the different fireball radii and the different number densities, and they are shown in the bottom right panel. During the calculation, we guarantee that the optical depth is less than the unity.

Basak \& Rao (2015) identified two thermal components fitted by the blackbodies in the X-ray spectrum of GRB 130925A.
They suggested a fast spine and a slow sheath layer in a structured jet to produce the blackbody spectra. Meanwhile, the blackbody radius was constrained to be $10^8-10^{10}$ cm. A similar radius number was also derived by \citet{bellm14}. As pointed out by \citet{bellm14}, the blackbody radius looks hard to be adopted in the jet structure model. Furthermore, when we use the blackbody radiation in the radius of $10^9$ cm to reproduce the thermal emission of the X-ray afterglow, a bulk Lorentz factor of 500 is required. This number seems relatively large according to normal GRB X-ray afterglow emission. If we consider the bremsstrahlung radiation in our model to reproduce the thermal feature in the GRB 130925A X-ray spectrum, we can abandon the above properties of the blackbody radiation that are not self-consistent.

If the thermal equilibrium is satisfied, the bremsstrahlung absorption coefficient is proportional to the term of $\nu^{-3}exp(-h\nu/kT)$. Because we consider the thermal properties in the X-ray band, even the number density has an order of $10^4~\rm{cm^{-3}}$, the absorption is negligible. The electron thermal equilibrium in our case is quite complex. In principle, the thermalization behind the shock is related to the fireball hydrodynamics, the energy transition from the external shock to the shocked material, and the energy allocation among different particles and magnetic field in the material. The dynamical timescale can be estimated by $t_{dyn}=\Delta H/c$, where $\Delta_H$ is the thickness of the shell. The bremsstrahlung cooling is estimated by $t_{ff}=(3/2)N_eKT/j(T)$, where $j(T)$ is the total emissivity \citep{rybicki1979}. We compare the two timescales and we find the condition of $t_{dyn}\ll t_{ff}$. This indicates that the electrons and the ions compressed by the shock can reach the thermal equilibrium before the energy dissipation by the bremsstrahlung radiation. Moreover, besides the shock acceleration that generates the nonthermal particles, the stochastic acceleration behind shocks generates the quasi-thermal particles that can be roughly described by the Maxwellian energy distribution \citep{sch89}. In this case, the thermal equilibrium condition can be satisfied.

\section{Discussion and Conclusions}
We have successfully applied the bremsstrahlung radiation to reproduce the thermal component in the X-ray afterglow of GRB 130925A observed by {\it Swift}-XRT and {\it NuSTAR} in this paper. In our scenario, number density of particles at a given fireball radius is essential to produce bremsstrahlung flux. The possibility of applying the bremsstrahlung radiation to explain the X-ray thermal component of GRB 130925A was also examined by \citet{bellm14}. In their work, the data were taken from {\it Swift}-XRT ($0.3-10$ keV), {\it Chandra} ($0.2-10$ keV) and {\it NuSTAR} ($3-30$ keV), and the bremsstrahlung model can well fit the spectrum.
Due to the large number density at the large fireball radius where the bremsstrahlung radiation occurred, \citet{bellm14} thought that the fitting was unphysical. In our calculation, with a bulk Lorentz factor of 10, after 1.8 days of the burst trigger, we obtain the fireball radius at about $2\times 10^{18}$ cm.
However, we note that Bellm et al. (2014) identified the thermal component below 10 keV within the energy range of 3-30 keV. While we consider the data from \citet{basak15b} and the thermal component by the bremsstrahlung radiation can be extended within the range of 20-80 keV, although the error bars of the data above 20 keV are very large. Our results do not argue against the expectation of Bellm et al. (2014) that the material in the emission region can be optical thin if high temperatures (about 50-80 keV) and a large radius (about $2\times 10^{18}$ cm) are adopted.

We agree that the large radius in our calculation is far beyond a typical range of normal GRB X-ray emission under the fireball framework.
However, we note that GRB 130925A is an ultra-long burst. The duration of the prompt emission is as long as about 20 ks, and the X-ray afterglow began from $2.0\times 10^4$ s after the trigger. A BSG/RSG population could be one possible progenitor class for the ultra-long GRB 130925A. These properties indicate that the X-ray afterglow of GRB 130925A may occur at a radius that is much larger than a normal one.
Furthermore, GRB dust destruction radius was limited to be about $10^{19}$ cm \citep{waxman00}. The radii given in our calculation are within this limit.
Therefore, it is still reasonable to consider the possibility that the bremsstrahlung radiation dominates the thermal component in the GRB 130925A X-ray afterglow at very large fireball radius.

BSG/RSG usually has strong stellar wind. \citet{piro14} identified the wind environment of GRB 130925A from the broadband spectrum.
It has been found that the environments of some GRBs were strongly polluted by metals and dust (e.g., Mao 2010; Perley et al. 2013; Wang \& Dai 2014; Perley et al. 2016).
\citet{schady15} measured the metallicity of $1.5Z_\odot$ from the optical spectrum of GRB 130925A, while GRBs usually have a metal-poor environment \citep{chris04,savaglio09}.
The absorption of $A_V=2.2$ that was obtained from the general host galaxy may be underestimated to
the real extinction. Here, we prefer the number of $A_V=7.7$ that was determined by the line of sight from the {\it Swfit}-XRT observation.
The dust destruction by the forward shock can provide enough ions and electrons for the bremsstrahlung radiation.
Moreover, heavy element nucleosynthesis in the condition of either a massive collapsar or compact binary merger was theoretically investigated \citep{janiuk14}. All of these heavy nuclei can be partly or fully ionized to provide ions and electrons for the bremsstrahlung radiation as well. The density profile of the surrounding medium can be also complex if RSG as a GRB progenitor ejects a convective envelope \citep{gendre13}. In this case, we also expect a large number density when we calculate the bremsstrahlung radiation.
In our calculation, we take the number density of $3.0\times 10^4~\rm{cm^{-3}}$ at the fireball radius of $2.0\times 10^{18}$ cm.
When we consider a RSG star (Fransson et al. 1996) or a Wolf-Rayet (WR) star (Chevalier \& Li 1999) as possible GRB progenitor, the number density of the wind-like environment can reach a number of $n\sim 10^6~\rm{cm^{-3}}$.
A larger number density at a smaller fireball radius or a smaller number density at a larger fireball radius can also be possible, on the condition that the optical depth of the bremsstrahlung radiation is smaller than the unity. On the other hand, we choose $Z=1$ in the calculation, and the numbers of $n_i$ and $n_e$ are the maximum ones. If the GRB environment has $Z>1$, we may have fewer ionized particles to reproduce the same bremsstrahlung flux if the ionization is strong. The much lower number density of $10^{-4}\le n\le 1.5~\rm{cm^{-3}}$ was given by \citet{evans14}. We note that this loose constraint has large uncertainties. \citet{horesh15} estimated a higher electron number density from a detailed radio analysis. The physics can be more complex if a binary system such as an ultra-long GRB progenitor is introduced \citep{stratta13}.

We suggest ultra-long GRB 130925A in a wind environment in this paper. In the calculation, we set the bulk Lorentz factor to be 10, and the material being close to the thermal equilibrium with a temperature of 50 keV is proposed. It seems that the results from the bremsstrahlung calculation are consistent with the observational data.
Some other ultra-long bursts, such as GRB 101225A \citep{campana11,campana11}, GRB 111209A \citep{stratta13}, and GRB 141121A \citep{cucchiara15}, were detected.
The thermal blackbody component was introduced to fit the X-ray spectra of GRB 101225A and GRB 111209A.
In general, we do not exclude the blackbody contribution to the afterglow emission in other ultra-long GRBs. In order to obtain significant bremsstrahlung radiation, large number densities of electrons and ions are required in the calculation. If the blackbody component is dominated in the X-ray thermal emission, or, if electrons and ions are not enough in the GRB environment, the bremsstrahlung radiation cannot be effective in the GRB thermal emission. In order to effectively detect the significant thermal emission of the X-ray afterglow that was produced by the bremsstrahlung radiation, detectors with the X-ray energy band beyond 10 keV are very helpful. We expect that {\it NuSTAR} and other similar X-ray telescopes
can perform more ultra-long GRB observations to further reveal the physics on the GRB progenitor and the GRB environment.

\acknowledgments
We are grateful to the referee for his/her helpful suggestions and comments.
J.-Y.L. acknowledges the financial support of the National Natural Science Foundation of China (grant 11303086).
J.M. is supported by the National Natural Science Foundation of China 11673062, the Hundred Talent Program of Chinese Academy of Sciences, the Major Program of Chinese Academy of Sciences (KJZD-EW-M06), and the Oversea Talent Program of Yunnan Province.

\begin{figure}
\includegraphics[scale=0.4]{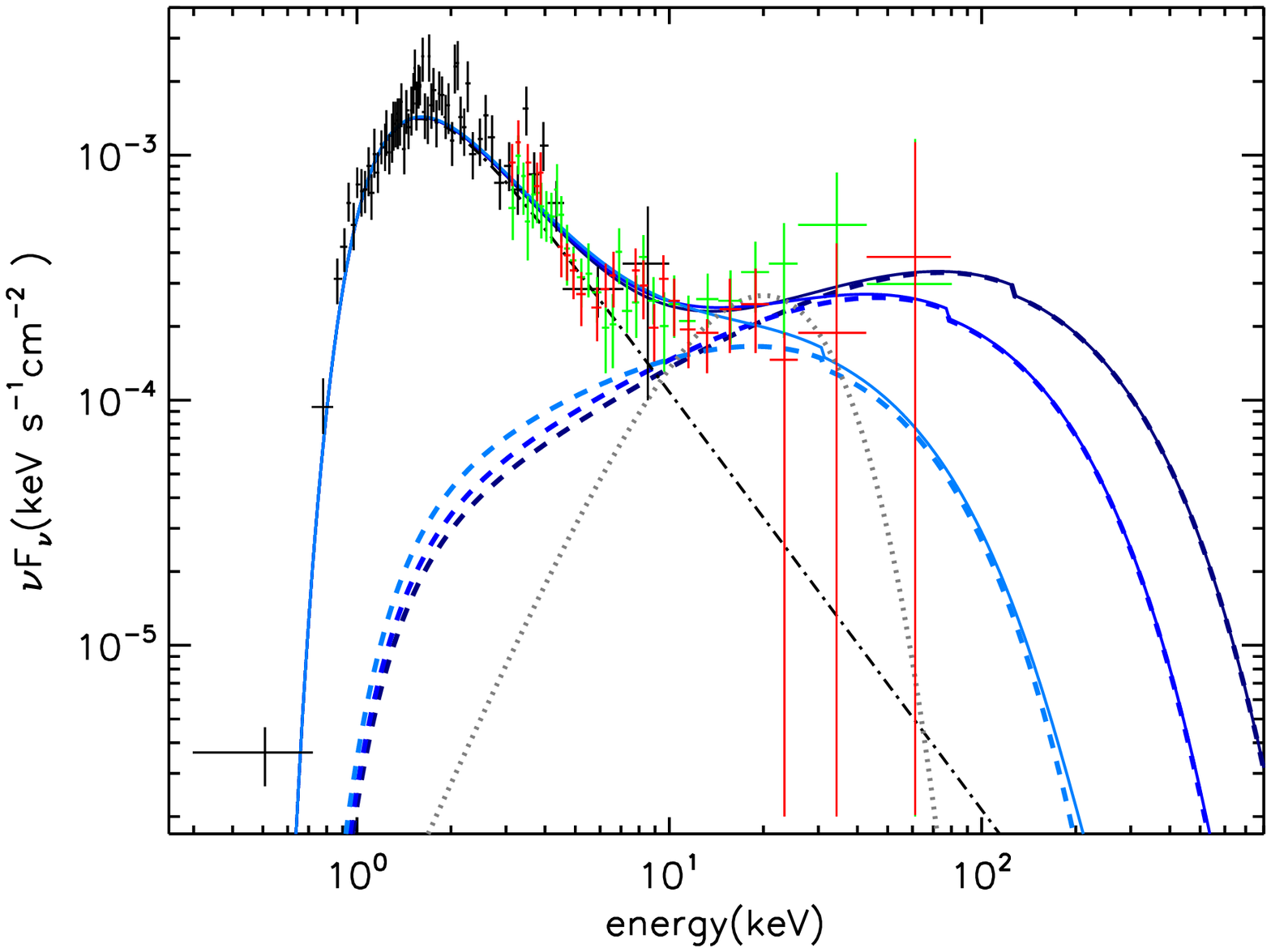}
\includegraphics[scale=0.4]{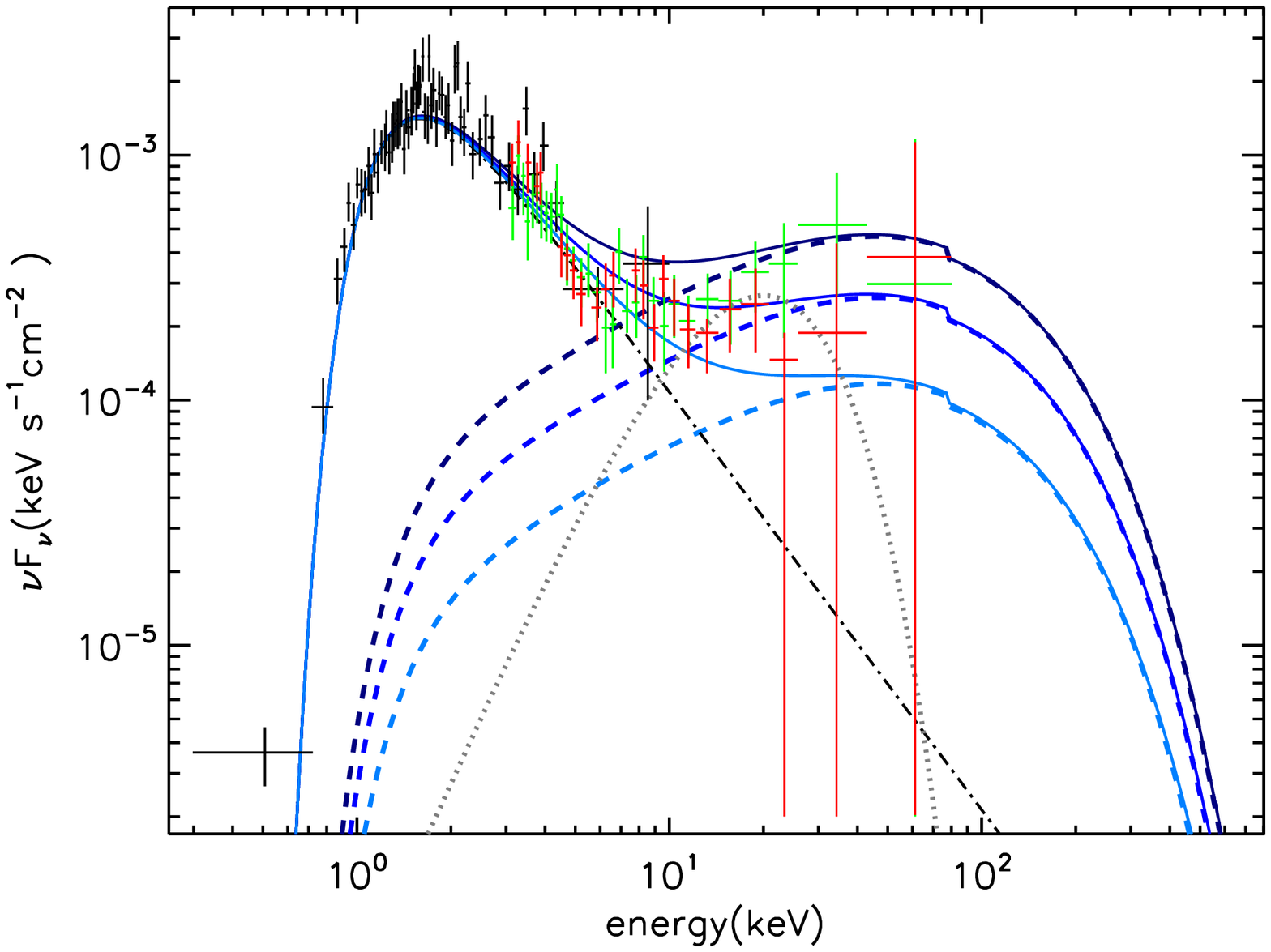}
\includegraphics[scale=0.4]{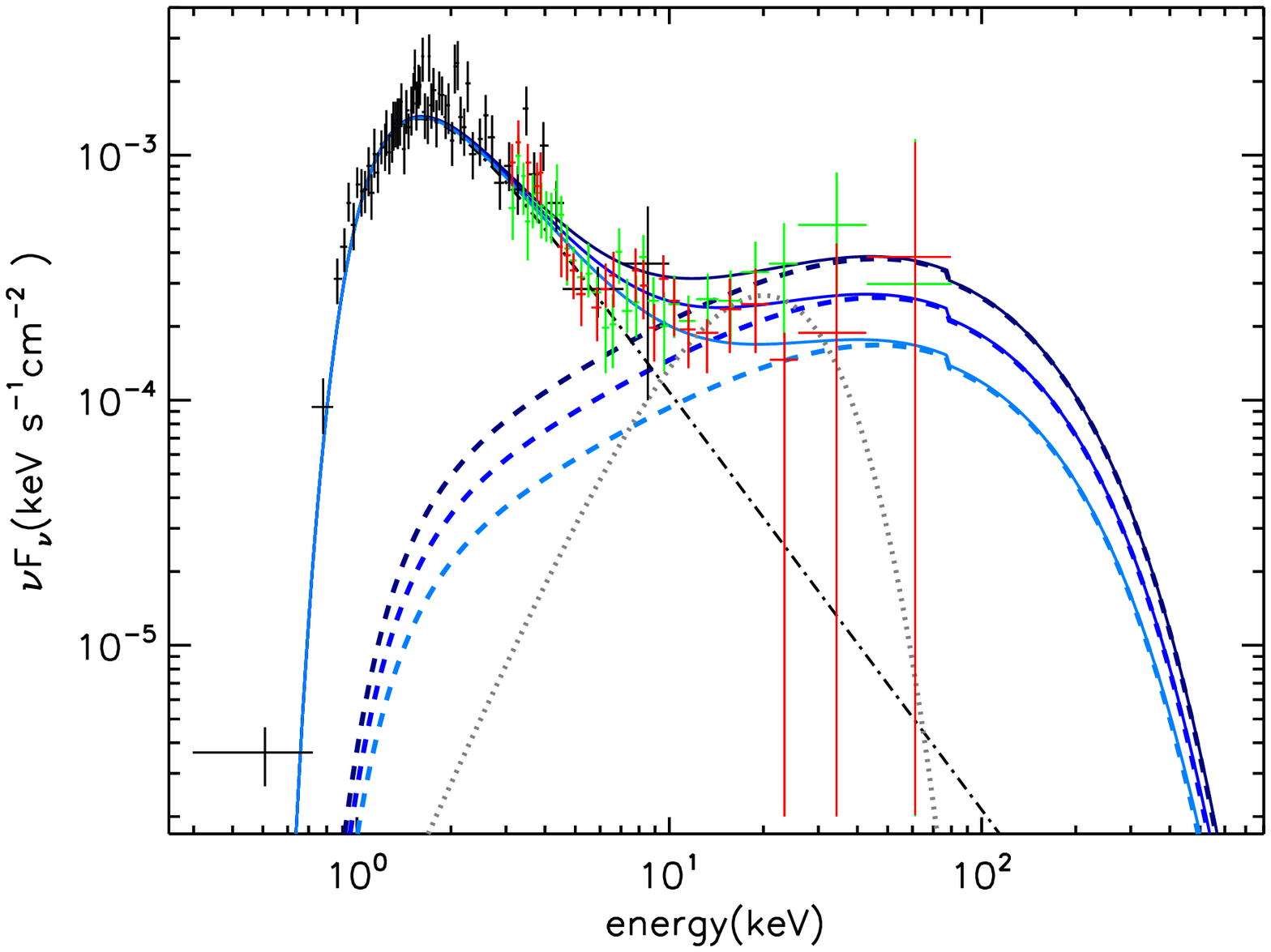}
\includegraphics[scale=0.4]{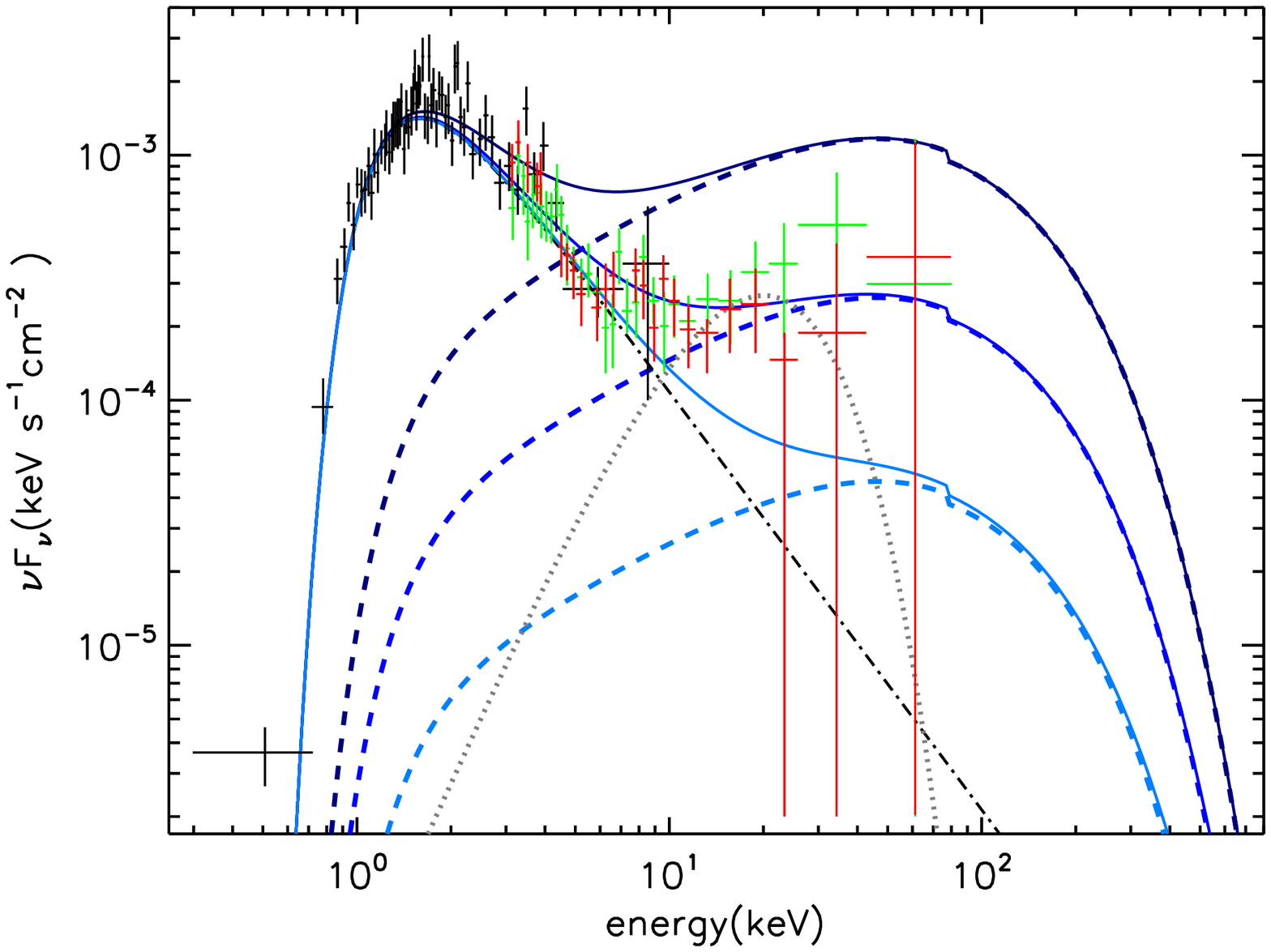}
\caption{Bremsstrahlung radiation for thermal emission of GRB 130925A X-ray afterglow.
Top left panel: the case of $n_e=n_i=3.0\times 10^4~\rm{cm^{-3}}$, $R=2.0\times 10^{18}$ cm, and $\Gamma=10.0$. Three solid lines with the colors of cyan, blue, and dark blue indicate the radiation of $kT_{\rm{obs}}=$20.0, 50.0, and 80.0 keV, respectively. Top right panel: the case of $kT_{\rm{obs}}=50.0$ keV, $R=2.0\times 10^{18}$ cm, and $\Gamma=10.0$. Three solid lines with the colors of cyan, blue, and dark blue indicate the radiation of $n_e=n_i=2.0\times 10^4, 3.0\times 10^4$, and $4.0\times 10^4~\rm cm^{-3}$, respectively. Bottom left panel: the case of $kT_{\rm{obs}}=50.0$ keV, $R=2.0\times 10^{18}$ cm, and $n_e=n_i=3.0\times 10^4~\rm{cm^{-3}}$. Three solid lines with the colors of cyan, blue, and dark blue indicate the radiation of $\Gamma$=8, 10, and 12, respectively. Bottom right panel: the case of $kT_{\rm{obs}}=50.0$ keV and $\Gamma=10.0$. Three solid lines with the colors of cyan, blue, and dark blue indicate the radiation of $R=2.0\times 10^{17}$ cm and $n_e=n_i=4.0\times 10^5~\rm{cm^{-3}}$, $R=2.0\times 10^{18}$ cm and $n_e=n_i=3.0\times 10^4~\rm{cm^{-3}}$, and $R=2.0\times 10^{19}$ cm and $n_e=n_i=2.0\times 10^3~\rm{cm^{-3}}$, respectively.
In all panels, the observational data of {\it Swift}-XRT (black) and {\it NuSTAR} (green and red) plotted by `$+$' are taken from \citet{basak15b}. The dotted-dashed line (black) indicates an absorbed power-law spectrum with a spectral index of 1.7. The absorption of the bremsstrahlung radiation is assumed to follow the same shape of the power-law spectrum.
The dotted line (grey) indicates a blackbody spectrum with an observed temperature of 5.1 keV.
\label{fig1}}
\end{figure}

\clearpage

\end{document}